\documentclass[aps,prl,preprint,amsfonts,superscriptaddress,showpacs,letterpaper]{revtex4}
\usepackage{graphicx}

\def \oli       {LiFePO$_4$}
\def \fepo      {FePO$_4$}
\def \olix      {Li$_x$FePO$_4$}
\def \ldapu     {GGA+U}

\begin{document}

\title{Configurational electronic entropy and the phase diagram of
mixed-valence oxides: the case of Li$_x$FePO$_4$}

\author{Fei Zhou}
\affiliation{Department of Physics, Massachusetts Institute of Technology, Cambridge, MA 02139, USA}
\author{Thomas Maxisch}
\affiliation{Department of Materials Science and Engineering, Massachusetts Institute of Technology, Cambridge, MA 02139, USA}
\author{Gerbrand Ceder}
\affiliation{Department of Materials Science and Engineering, Massachusetts Institute of Technology, Cambridge, MA 02139, USA}

\date{\today}
\pacs{64.75.+g, 65.40.Gr, 82.47.Aa, 71.15.Mb}

\begin{abstract}
We demonstrate that configurational electronic entropy, previously neglected, in {\it ab initio}
thermodynamics of materials can qualitatively modify the finite-temperature phase stability
of mixed-valence oxides. While transformations from low-T ordered or immiscible states are almost
always driven by configurational disorder (i.e.\ random occupation of lattice sites
by multiple species), in FePO$_4$--LiFePO$_4$ the formation of a solid solution is almost entirely driven by electronic, rather than ionic
configurational entropy. We argue that such an electronic entropic mechanism 
may be relevant to most other mixed-valence systems.
\end{abstract}
\maketitle

First-principles prediction of a crystalline material's phase
diagram based on the density functional theory (DFT) is a
prime example of the achievement of modern solid state physics \cite{Fontaine1994}.
A pure DFT approach is applicable to zero-temperature
(zero-T). To study finite-T phase stability, one has to
identify carefully all the excitations and degrees of freedom
involved in creating entropy. Typically in alloy theory the focus is on the configurational disorder
(substitution of different elements or vacancies (V))
while the electronic degrees of freedom are, in the spirit of the
adiabatic approximation, integrated out \cite{Fontaine1994, Ceder1993CMS} (phonon contributions may give quantitative
corrections \cite{Walle2002RMP, Asta1993PRB, Wolverton2001PRL},
especially in systems with
exotic electron-phonon coupling \cite{Manley2006PRL}, but they are relatively
composition insensitive and will not be discussed here). For
example, many phase diagrams can be
satisfactorily reproduced by considering the configurational entropy of
two elements \cite{Fontaine1994} or element and vacancy \cite{Van1998PRB}.
Electronic entropy is usually thought of as a small quantitative correction and can be
calculated from the band structure:
\cite{Wolverton1995PRB-First-Principles, Nicholson1994PRB}:
\begin{equation}
S_{{\rm e}}^{{\rm band}} = - k_B \int n (f
\ln f + (1-f) \ln(1-f)) dE,
\end{equation}
where $n$ and $f$ are the density of states and Fermi distribution
function, respectively. 
Only electrons within $\sim k_B T$ to the Fermi level
participate in the excitations, so $S_{{\rm e}}^{{\rm band}}$ is usually small.
A different type of electronic entropy could arise if electrons/holes (e/h)
are localized and contribute to the total entropy in the same fashion
as the ordering of atoms.
One would expect such configurational electronic entropy to be particularly important in
mixed-valence transition metal oxides.
Many technologically important materials, such as doped manganites,
high-T superconductors, Na- and Li-metal oxides, and mixed conductors,
fall in this category.
Little is known about the contribution of localized e/h to finite-T phase stability, though
previous evidence exist in
doped superconductors \cite{Schleger1994PRB, Tetot1999PRB} and perovskites \cite{Lankhorst1997SSI}
that a configurational electronic entropy term (assuming random e/h distribution):
\begin{equation} \label{eq:PD-loc-entropy}
S_{{\rm e}}^{{\rm loc,rand}}=- k_B \left[x\ln x + (1-x) \ln (1-x)\right],
\end{equation}
helps explain the entropy of oxidation/reduction.
In Eq.~\ref{eq:PD-loc-entropy} $x$ is the concentration of localized electrons or holes.
While $S_{{\rm e}}^{{\rm loc,rand}}$ can potentially
be as significant as the configurational entropy of ions.
there exists currently no clear demonstration that electronic entropy can qualitatively modify
finite-T phase diagram.

In this letter we investigate the effects of configuration-dependent electronic entropy.
We go beyond a random model such as Eq.~\ref{eq:PD-loc-entropy} and sample electron configurations
explicitly.
We focus on the \olix\ system.
While its high intrinsic Li$^+$ mobility
makes it of interest as the
next-generation cathode for rechargeable Li batteries \cite{Padhi1997JES-Phospho-olivines}, it is also crucial to ensure
good phase equilibration, even at room temperature (RT). So \olix\ is a good system to benchmark theory against.
We show that excellent agreement with the experimental
phase diagram can only be achieved by taking into account configurational electronic entropy,
and qualitative discrepancies occur if the electron degree of freedom
is ignored.

\begin{figure}
\includegraphics[width=0.99 \linewidth]{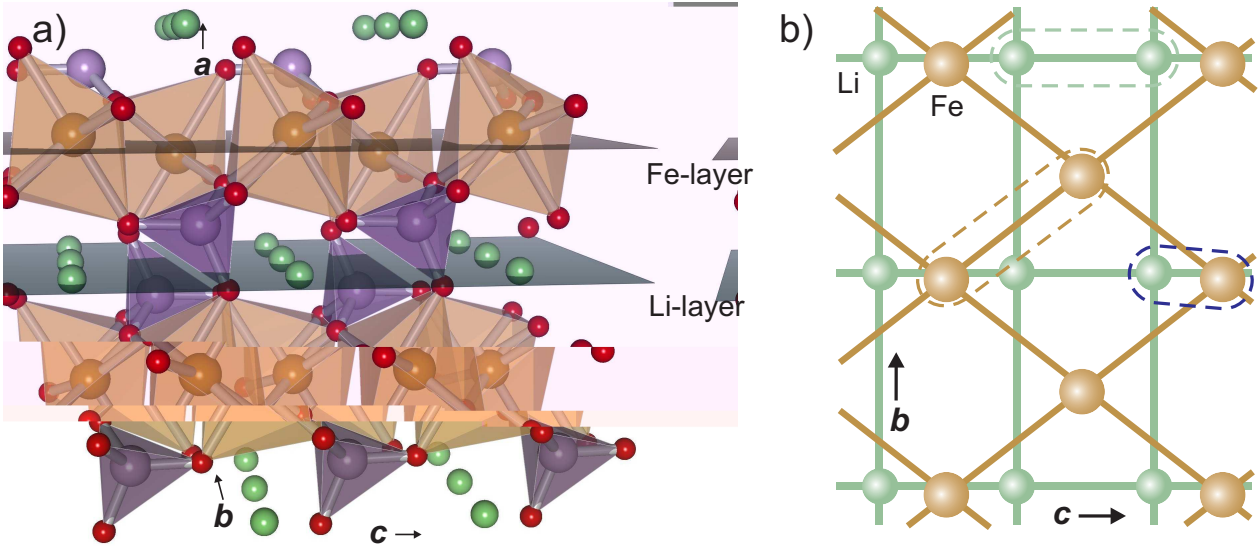}
\caption{The \oli\ structure shown with:
{\bf a}) PO$_4$ (purple) and FeO$_6$ (brown) polyhedra as well as Li atoms (green)
{\bf b}) adjacent
layers on Li and Fe sub-lattices, projected along axis $a$, with nearest-neighbor (NN) inter- and
intra-lattice pairs highlighted.} \label{fig:PD-structure}
\end{figure}
\begin{figure}
\includegraphics[width=0.8 \linewidth]{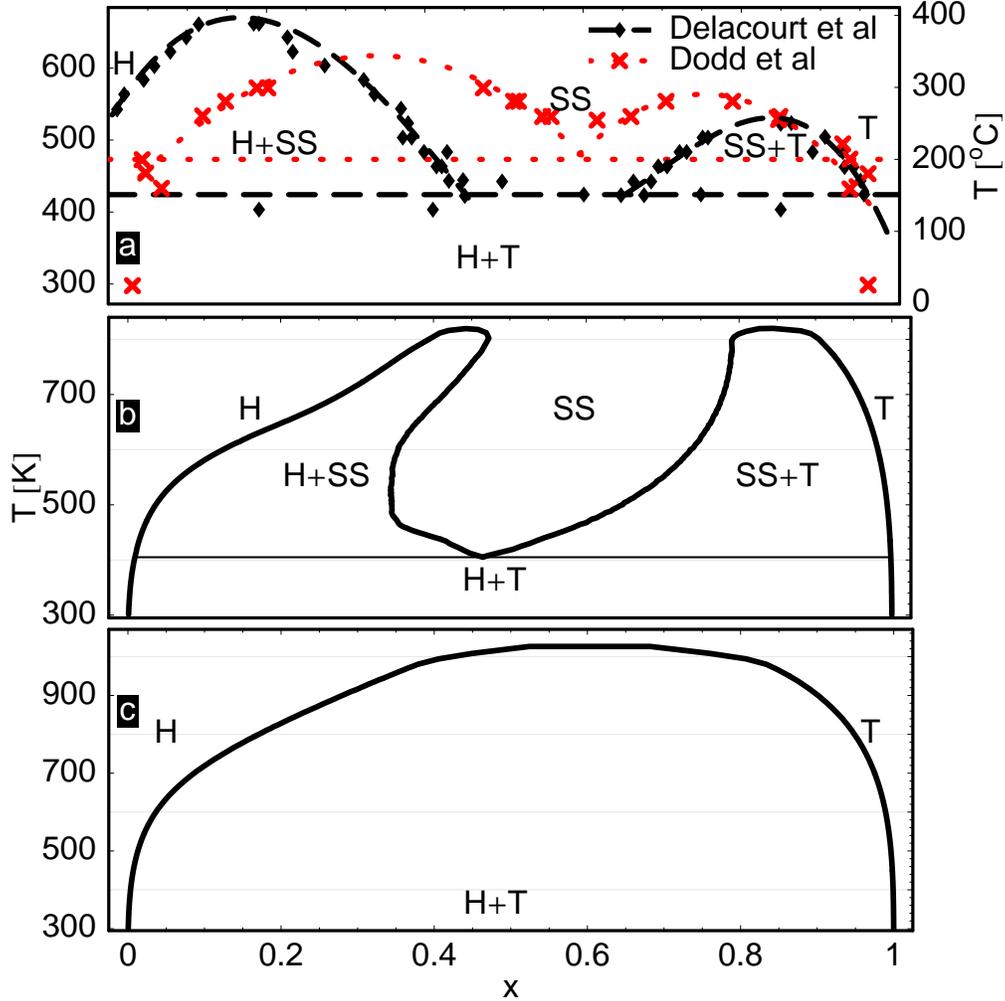}
\caption{\olix\ phase diagram.
{\bf a}) experimental phase boundary data taken from Delacourt et al \cite{Delacourt2005NM}
and from Dodd et al \cite{Dodd2006ESSL};
{\bf b}) calculated with both Li and electron degrees of freedom and 
{\bf c}) with explicit Li only.
} \label{fig:PD-boundary}
\end{figure}

\oli\ has an olivine-type structure with an orthorhombic unit cell.
Li removal at RT occurs through a miscibility gap
between triphylite (T)  \oli\ and heterosite (H) \fepo\ \cite{Padhi1997JES-Phospho-olivines}
with both phases having a  very limited amount of solubility
(vacancies + holes (Fe$^{3+}$) in T and Li$^+$ ions + electrons (Fe$^{2+}$) in H) \cite{Yamada2005ESSL-Phase}.
Recent higher-T investigations of the \olix\ phase diagram by Delacourt et al
\cite{Delacourt2005NM} and by Dodd et al \cite{Dodd2006ESSL}
confirm the low-T immiscibility, but also find an unusual
eutectoid point at 150$^\circ$C
\cite{Delacourt2005NM} or 200$^\circ$C \cite{Dodd2006ESSL} where the
solid solution (SS) phase emerges around $x \approx 0.45 - 0.65$.
Above 300--400 $^\circ$C SS dominates all compositions (Fig.~\ref{fig:PD-boundary}a).

This phase diagram is quite unexpected from a
theoretical point of view. First, why does the system phase
separate at all at low-T? In a simplified picture of
a generic oxide Li$_x$MO$_n$, the Li$^+$ ions repel each
other due to electrostatics so that ordered intermediate compounds
are energetically favorable over phase separation, i.e. segregation of Li$^+$ (vacancies) into Li-rich
(deficient) regions. This is indeed the case
in many other materials, in which mobile ions and vacancies coexist, e.g. Li$_x$CoO$_2$,
Li$_x$NiO$_2$, Na$_x$CoO$_2$ \cite{Van1998PRB, Terasaki1997PRB, Dompablo2002PRB}.
Secondly, what is the origin of the complex high-T behavior?
Transitions from a two-phase coexistence state to a solid solution
are typically driven by the configurational entropy of the ions in the SS, with a maximum transition T near equiatomic A/B composition.
The experimentally established phase diagram, shown in Fig.~\ref{fig:PD-boundary}a, is unlikely to come
from such ionic configurational entropy unless the effective Li-V interactions are
unusually strongly composition dependent.
We demonstrate that the topology arises from electron degrees of freedom
which stabilizes the SS near $x \approx 0.5$.

We study the \olix\ phase diagram by Monte Carlo simulations based on a coupled cluster expansion \cite{Sanchez1984PA,Tepesch1995PRL},
which is a Hamiltonian that explicitly describes the dependence of the energy on the
arrangement of Li$^+$/V  and Fe$^{2+}$/Fe$^{3+}$, i.e.\ both ionic and
electronic degrees of freedom. In \olix\ the Li$^+$ ions and vacant sites sit
on an orthorhombic lattice, of which one layer is shown in Fig.~\ref{fig:PD-structure}b
(large green points). On each side of this Li layer is a plane of Fe sites
(only one plane shown in small brown points).
Representing with $\lambda_i\ =\pm 1$  occupation of site $i$
by a Li$^+$ or vacancy  and with $\epsilon_a =\pm 1$ the presence of Fe$^{2+}$ (electron)
or Fe$^{3+}$ (hole) on site $a$, the energy can be expanded without loss of generality in
polynomials of these occupation variables \cite{Sanchez1984PA,Tepesch1995PRL}:
\begin{equation} \label{eq:PD-CE}
E[\vec{\lambda}, \vec{\epsilon}]= J_{\emptyset} +  J_i \lambda_i +
 J_{ij} \lambda_i \lambda_j +
 J_{ia} \lambda_i \epsilon_a + J_{ab} \epsilon_a \epsilon_b + \dots
\end{equation}
The expansion coefficients $J$ are called  effective cluster interactions (ECI), essentially
coupling constants in a generalized Ising model.
In its untruncated form, Eq.~\ref{eq:PD-CE} is exact and includes
all multi-body terms within one sub-lattice (Li/V or e/h) and between
sub-lattices though some truncation takes place in practice.
To parameterize Eq.~\ref{eq:PD-CE} we have performed \ldapu\ calculations for 245 \olix\ ($0\leq x\leq1$)
configurations with super-cells of up to 32
formula units using parameter $U-J=4.3$ eV
\cite{Zhou2004PRB-First-principles} and other settings in \cite{Zhou2004PRB-Phase, Zhou2004PRB-First-principles}.
For each Li/V configuration, usually more than one e/h configuration were considered.
The \ldapu\ \cite{Anisimov1991PRB-Band} approach is essential to properly localize electronic
states (polarons) in this material \cite{Zhou2004PRB-Phase, Maxisch2006PRB}.
Removal of the self-interaction
through proper treatment of the on-site electron correlation of localized $d$-electrons in \ldapu\ has
previously shown to accurately reproduce the band gap \cite{Zhou2004SSC}, lithium insertion voltage
\cite{Zhou2004PRB-First-principles, Zhou2004EC,Le2005PM} and low-T immiscibility \cite{Zhou2004PRB-Phase}, unlike
uncorrected LDA or GGA which incorrectly predict stable intermediate \olix\ compounds \cite{Zhou2004PRB-Phase}.
Ferromagnetic high-spin Fe ions are assumed.
At RT (Li)\fepo\ is paramagnetic \cite{Rousse2003CM} and energetic effects of magnetic ordering
are small \cite{Zhou2004PRB-Phase},
so the spin entropy ($\approx k_B [x\ln 5+(1-x)\ln 6]$) is linear
in $x$ at RT, therefore negligible in phase diagram calculations.

\begin{figure}
\includegraphics[width=0.80 \linewidth]{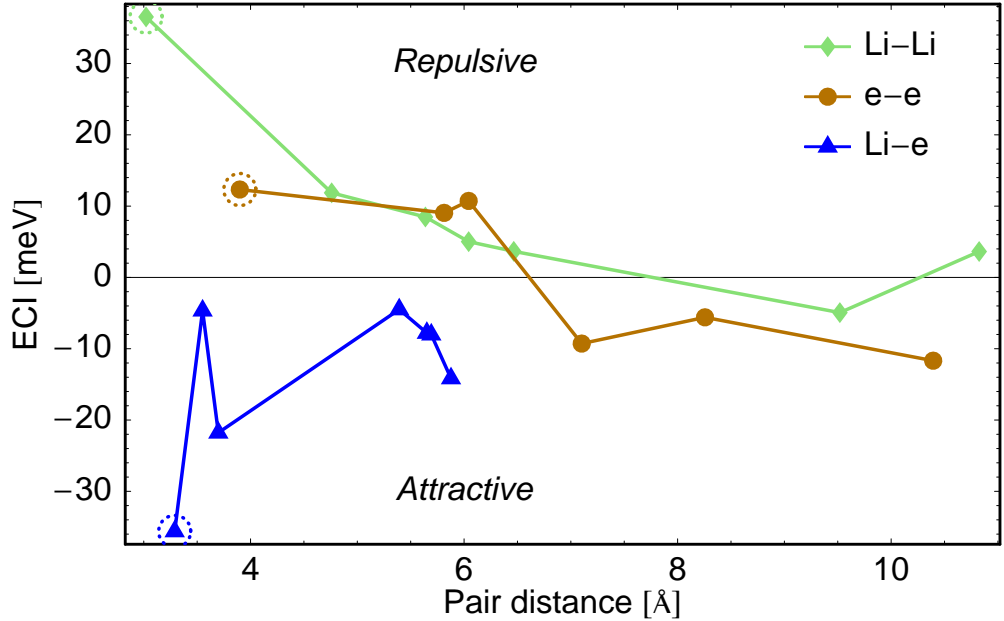}
\caption{Pair ECI vs.\ site distance (measured from the sites' ideal coordinates in \oli).
The circled points correspond to NN Li-Li, e-e and Li-e pairs in Fig.~\ref{fig:PD-structure}b.
} \label{fig:PD-paireci}
\end{figure}
Our cluster expansion model consists of 29 distinct ECIs:
the constant and the point terms with no effect on the phase diagram;
7 small triplet terms, which mainly represent slight
asymmetry between \fepo\ and \oli; and most significantly 20 pair
interactions shown in Fig.~\ref{fig:PD-paireci}.
Note that these are {\em effective} interactions including
the effects of many physical factors: electrostatics, screening, relaxation,
covalency, etc.
The Li-Li ECI (diamond) is largest for nearest-neighbor (NN) Li$^+$ ions, which repel each other
strongly for electrostatic reasons. As the pairs are separated further,
the repulsion is screened considerably. The small negative
$J_{\mathrm{Li-Li}}$ at large distance indicates some mediation of the effective
interactions by lattice distortions. Roughly the same trend is
observed for $J_{\mathrm{e-e}}$. On the
contrary, the Li-e inter-lattice ECIs are strong short-range
attractions that generally become weaker at longer distance. The
trend in the three curves is not monotonic, since the ECIs contain
complex lattice factors beyond isotropic electrostatics. The low-T
phase separation can be explained by considering the dominating
short-range terms. The Li$^+$ ions repel each other and so do
electrons, while Li-e attractions compete to bind
them together: if Li$^+$ ions stay together then the e$^-$ can bind to
more of them. The Li-e attractions prevail partly because of the host's
geometry:
the multiplicity of the NN Li-e ECI, the strongest attraction, is two per formula unit, while
that of NN Li-Li ECI, the strongest repulsion, is one (see Fig.\ \ref{fig:PD-structure}).
We therefore conclude that phase separation in \olix\ is mainly driven by
Li-e attractions in competition with Li-Li and e-e repulsions.
This is fundamentally different from a system where the
electronic mixed valence is delocalized, as in metallic Li$_x$CoO$_2$ \cite{Van1998PRB}, thereby making the Li-e coupling independent of
the Li/V distribution.

Monte Carlo (MC) simulations combining canonical (interchanging Li/V or e/h pairs)
and grand canonical (interchanging Li+e together into V+h and vice versa) steps
are carried out  on a $6 \times 12 \times 12$ super-cell,
resulting in the phase diagram in Fig.\ \ref{fig:PD-boundary}b.
Phase boundaries were obtained with free energy integration.
In excellent agreement with \cite{Delacourt2005NM,Dodd2006ESSL},
the calculated phase diagram features a miscibility gap between \fepo
\ and \oli, and
an unusual eutectoid transition to the solid solution phase.
The eutectoid temperature is only 20--70 K off from \cite{Delacourt2005NM,Dodd2006ESSL}, and the congruent temperatures
are about 100--150 K off.
We predict the enthalpy of mixing at the eutectoid point to be 8.6
meV/formula unit, consistent with the measured lower limit 700 J/mol= 7.3 meV/formula unit
for an $x=0.47$ sample \cite{Dodd2006ESSL}.

To understand better which physics determines the shape of Fig.~\ref{fig:PD-boundary}b,
we have also performed calculations in the more ``traditional'' way, i.e. to consider only the
Li/V ordering as the entropy generating mechanism,
assuming electrons always occupy the lowest energy state for each Li/V configuration.
The calculated phase diagram (Fig.~\ref{fig:PD-boundary}c) shows a simple two-phase
region, qualitatively different from experiment
but similar to typical immiscible systems.
The striking difference between Fig.~\ref{fig:PD-boundary}b and \ref{fig:PD-boundary}c
points to the {\em crucial importance of explicitly treating the electron degrees of freedom in excitations
and finite-T thermodynamics of these mixed-valence systems}.

\begin{figure}
\includegraphics[width=0.78 \linewidth]{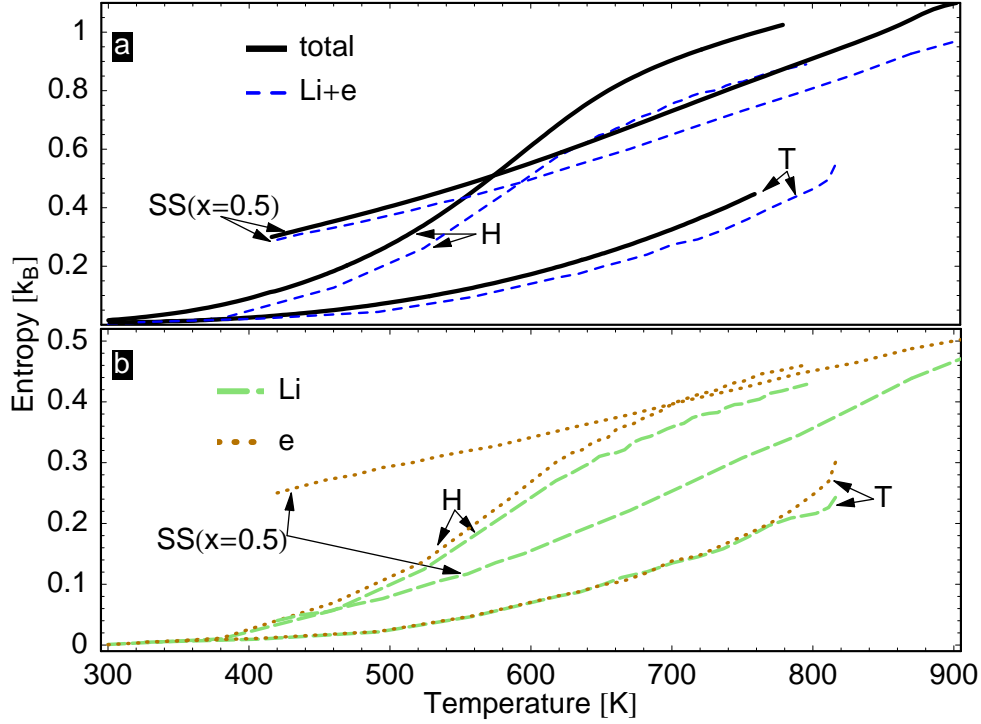}
\caption{Configurational entropy per formula unit.
{\bf a}) total entropy and the sum $S'_{{\rm Li}} + S'_{{\rm e}}$;
{\bf b}) separate conditional entropy $S'_{{\rm Li}}$ and $S'_{{\rm e}}$.
} \label{fig:PD-entropy}
\end{figure}
%
A deeper analysis of the phase diagram in Fig.~\ref{fig:PD-boundary}b requires investigation of
the entropy driving the phase transition. The total (joint) configurational
entropy $S(\mathrm{Li,e})$ of the electronic+ionic system can be calculated through free energy integration. 
To partition the entropy into ionic and electronic contributions, we
note that
\begin{equation} \label{eq:PD-entropy}
S(\mathrm{Li,e}) = S'(\mathrm{Li})+ S'(\mathrm{e}) + I(\mathrm{Li,e}),
\end{equation}
where $I$ is the mutual information of the two degrees of freedom, and
$S'(X)\equiv S(X|Y)= \sum_y P(y)S(X|y)$ is the conditional entropy from the X (Li or electron) degree of
freedom, i.e.\ the entropy contribution of X with fixed Y, thermal averaged over the marginal distribution
$P(Y)$. $S'(X)$ measures how random X can be when Y is fixed. If X and Y are independent, $S'$ is
exactly the entropy contribution from one degree of freedom. 
We use $S'$ to compare different entropy contributions.
In Fig.~\ref{fig:PD-entropy} we show the total and separate
entropy along the solubility limits of the H and T phases (leftmost and rightmost
phase boundaries in Fig.~\ref{fig:PD-boundary}b, respectively), as well
as along $x=0.5$ in SS.
At low-T the total entropy (bold lines in Fig.~\ref{fig:PD-entropy}a) is small, slightly larger in
H than in T.
The solid solution phase is far from random: (1)
when it first appears at the eutectoid point, its entropy is a mere 0.3 $k_B$, (2)
the total entropy of the H phase exceeds that of SS above
about 570 K even though its Li content is lower, and (3)
up to 900 K, well above the congruent points, the total entropy 1.1 $k_B$ of SS($x=0.5$) is still smaller than
(complete random) $2 S^{\mathrm{loc,rand}}_\mathrm{e}(0.5)=1.39\ k_B$.
The difference between $S(\mathrm{Li,e})$ and $S'_{{\rm Li}} + S'_{{\rm e}}$ (thin dashed curve of Fig.~\ref{fig:PD-entropy}a)
is the mutual information $I(\mathrm{Li,e})$, indicating how correlated the two degrees of freedom are.
Fig.~\ref{fig:PD-entropy}b shows separate $S'_{{\rm Li}}$ and $S'_{{\rm e}}$ in
dashed and dotted curves, respectively. It is noteworthy that in all but the T branches
$S'_{{\rm e}}$ is noticeably larger than $S'_{{\rm Li}}$;
$S'_{{\rm e}}$ dominates the SS phase and contributes
much more than $S'_{{\rm Li}}$.
At the eutectoid point the mixing entropy driving the transition into SS is overwhelmingly electronic: 0.19 $k_B$ from $S'_{{\rm e}}$
vs.\ 0.05 $k_B$ from $S'_{{\rm Li}}$.
A qualitative explanation for the larger $S'_{{\rm e}}$ is that the leading $J_\mathrm{e-e}$
terms are weaker than the leading $J_\mathrm{Li-Li}$, and the electron excitation spectrum
at a fixed Li configuration is lower in energy than the opposite.
We therefore conclude, to the extent $S'$ represents a separate entropy,
that the electron degree of freedom contributes substantially more than Li ions to disordering of the system,
and that the formation of the solid solution state is driven by e/h disorder.
To our knowledge, no other examples of electronic entropy-driven solid solution have been identified, though electronic
entropy driven modification of ordering interactions through band entropy has been proposed for Ni$_3$V \cite{Johnson2000PRB}.


Beyond \oli, our approach and results may help our understanding of other mixed-valence transition metal oxides
with localized electrons. 
In oxides both electron localization and delocalization can occur. For example, a system
such as Li$_x$CoO$_2$ is metallic for $x<0.9$ \cite{Menetrier1999JMC} and explicit e/h entropy is less crucial.
LDA and GGA in which mixed valence states are delocalized will be an adequate treatment
for such system \cite{Van1998PRB}. On the other hand materials in which carriers localize  require
more careful treatment both for their energy calculation (e.g.\ in \ldapu, SIC methods or DMFT \cite{Georges1996RMP}),
and for their contribution of the electronic degree of
freedom to the entropy as demonstrated in the present work.
An even more
complicated situation arises in materials where electrons can be exchanged
between localized and delocalized states, as in Ce \cite{Svane1996PRB-Electronic}.
It should be noted that in our MC simulations, e/h are treated as classical particles
(but not in the DFT energy calculations).
If hopping becomes so fast that electron wavefunctions overlap, the notion of localized electrons becomes meaningless,
and it becomes difficult to enumerate the eigenstates over which to sum excitations, until
one reaches the nearly free-electron limit where the band picture is applicable.
It is up to further investigation
to establish quantitative effects of the localized electron degree of freedom
in thermodynamics of other transition metal oxides.

FZ would like to thank A.~Van der Ven for his help in computation.
This work is supported by DOE under contract
DE-FG02-96ER45571 and by the NSF MRSEC program under
contract DMR-0213282.

\end{document}